\documentclass[runningheads]{llncs}

\usepackage{listings}
\usepackage{color}
\usepackage{url}
\usepackage{hyperref}
\usepackage{tikz}
\usepackage{balance}
\usepackage{multirow}
\usepackage{graphicx}
\usepackage{wrapfig,lipsum,booktabs}
\usepackage{comment}
\usepackage{subfigure}

\definecolor{dkgreen}{rgb}{0,0.6,0}
\definecolor{gray}{rgb}{0.5,0.5,0.5}
\definecolor{mauve}{rgb}{0.58,0,0.82}

\newcommand{\MySection}[1]{
  \vspace{-2.5ex}
  \section{#1}
  \vspace{-2.0ex}
}
\newcommand{\MySubsection}[1]{
  \vspace{-2.5ex}
  \subsection{#1}
  \vspace{-1.5ex}
}

\usepackage{pifont}
\newcommand{\cmark}{\ding{52}}
\newcommand{\xmark}{\ding{53}}

\newcommand{\mytwocolfigcustom}[7]{
\begin{figure}[htbp]
  \begin{center}
      \mbox{
      \hspace{-0.3\columnsep}
      \subfigure[#1]
      {
          \includegraphics[width=0.52\columnwidth]{#2}
          \label{#3}
      }
      \hspace{-0.3\columnsep}
      \subfigure[#4]
      {
          \includegraphics[width=0.44\columnwidth]{#5}
          \label{#6}
      }
      }
      \vspace*{-1.0\baselineskip}
      \caption{#7}
  \end{center}
\vspace*{-1\baselineskip}
\end{figure}
}

\newcommand{\mytwocolfig}[7]{
\begin{figure}[htbp]
\vspace{-4.0ex}
  \begin{center}
      \mbox{
      \hspace{-0.3\columnsep}
      \subfigure[#1]
      {
          \includegraphics[width=0.48\columnwidth]{#2}
          \label{#3}
      }
      \hspace{-0.3\columnsep}
      \subfigure[#4]
      {
          \includegraphics[width=0.48\columnwidth]{#5}
          \label{#6}
      }
      }
      \vspace*{-1.0\baselineskip}
      \caption{#7}
      \vspace{-4.0ex}
  \end{center}
\vspace*{-1\baselineskip}
\end{figure}
}

\newcommand{\mythreecolfig}[9]{
\begin{figure}[htbp]
\vspace{-4.0ex}
  \begin{center}
      \mbox{
      \hspace{-0.31\columnsep}
      \subfigure[#1]
      {
          \includegraphics[width=0.31\columnwidth]{#2}
          \label{#3}
      }
      \hspace{-0.31\columnsep}
      \subfigure[#4]
      {
          \includegraphics[width=0.31\columnwidth]{#5}
          \label{#6}
      }
      \hspace{-0.31\columnsep}
      \subfigure[#7]
      {
          \includegraphics[width=0.31\columnwidth]{#8}
          \label{#9}
      }
      }
      \vspace*{-1.0\baselineskip}
      \caption{Testing the Correctness of HyPar-Flow using Different Models}
      \vspace{-6.0ex}
  \end{center}
\vspace*{-1\baselineskip}
\end{figure}
}

\begin{document}
\title{HyPar-Flow: Exploiting MPI and Keras for Scalable \underline{Hy}brid-\underline{Par}allel DNN Training with Tensor\underline{Flow}}

\titlerunning{HyPar-Flow (Accepted to be presented at ISC '20)}

\authorrunning{A. A. Awan, A. Jain, Q. Anthony, H. Subramoni, and DK Panda}

\author{Ammar Ahmad Awan \and Arpan Jain \and Quentin Anthony \and Hari Subramoni \and Dhabaleswar K. Panda}

\institute{The Ohio State University, Columbus, OH 43210, USA\\  \email{\{awan.10, jain.575, anthony.301, subramoni.1, panda.2\}@osu.edu}}

%\author{}

\maketitle

\begin{abstract}
To reduce the training time of large-scale Deep Neural Networks (DNNs), scientists have started to explore parallelization strategies like data-parallelism, model-parallelism, and hybrid-parallelism. While data-parallelism has been extensively studied and developed, several problems exist in realizing model-parallelism and hybrid-parallelism efficiently. Four major problems we focus on are: 1) defining a notion of a distributed model across processes, 2) implementing forward/back-propagation across process boundaries that requires explicit communication, 3) obtaining parallel speedup on an inherently sequential task, and 4) achieving scalability without losing out on a model's accuracy. To address these problems, we create \textbf{HyPar-Flow} --- 
a model-size/-type agnostic, scalable, practical, and user-transparent system for hybrid-parallel training by exploiting MPI, Keras, and TensorFlow. HyPar-Flow provides a single API that can be used to perform data, model, and hybrid parallel training of any Keras model at scale. We create an internal distributed representation of the user-provided Keras model, utilize TF's Eager execution features for distributed forward/back-propagation across processes, exploit pipelining to improve performance and leverage efficient MPI primitives for scalable communication. Between model partitions, we use \textit{send} and \textit{recv} to exchange layer-data/partial-errors while \textit{allreduce} is used to accumulate/average gradients across model replicas. Beyond the design and implementation of HyPar-Flow, we also provide comprehensive correctness and performance results on three state-of-the-art HPC systems including TACC \textit{Frontera} (\#5 on Top500.org). For ResNet-1001, an ultra-deep model, HyPar-Flow provides: 1) Up to 1.6$\times$ speedup over Horovod-based data-parallel training, 2) 110$\times$ speedup over single-node on 128 Stampede2 nodes, and 3) 481$\times$ speedup over single-node on 512 Frontera nodes.
\end{abstract}

\keywords{Hybrid Parallelism, Model Parallelism, Keras, TensorFlow, MPI, Eager Execution, Deep Learning}

\MySection{Introduction and Motivation}
\label{sec:intro}

Recent advances in Machine/Deep Learning (ML/DL) have triggered key success stories in many application domains like Computer Vision, Speech Comprehension and Recognition, and Natural Language Processing. Large-scale Deep Neural Networks (DNNs) are at the core of these state-of-the-art AI technologies and have been the primary drivers of this success. However, training of DNNs is a compute-intensive task that can take weeks or months to achieve state-of-the-art prediction capabilities (\textit{accuracy}). These requirements have led researchers to resort to a simple but powerful approach called \textit{data-parallelism} to achieve shorter training times. Various research studies~\cite{scaffe,goyal2018} have addressed performance improvements for data-parallel training. As a result, production-grade ML/DL software like TensorFlow and PyTorch also provide good support for data-parallelism. 

\begin{figure}[htbp]
\centering
    \vspace{-4.0ex}
    \includegraphics[width=0.7\columnwidth]{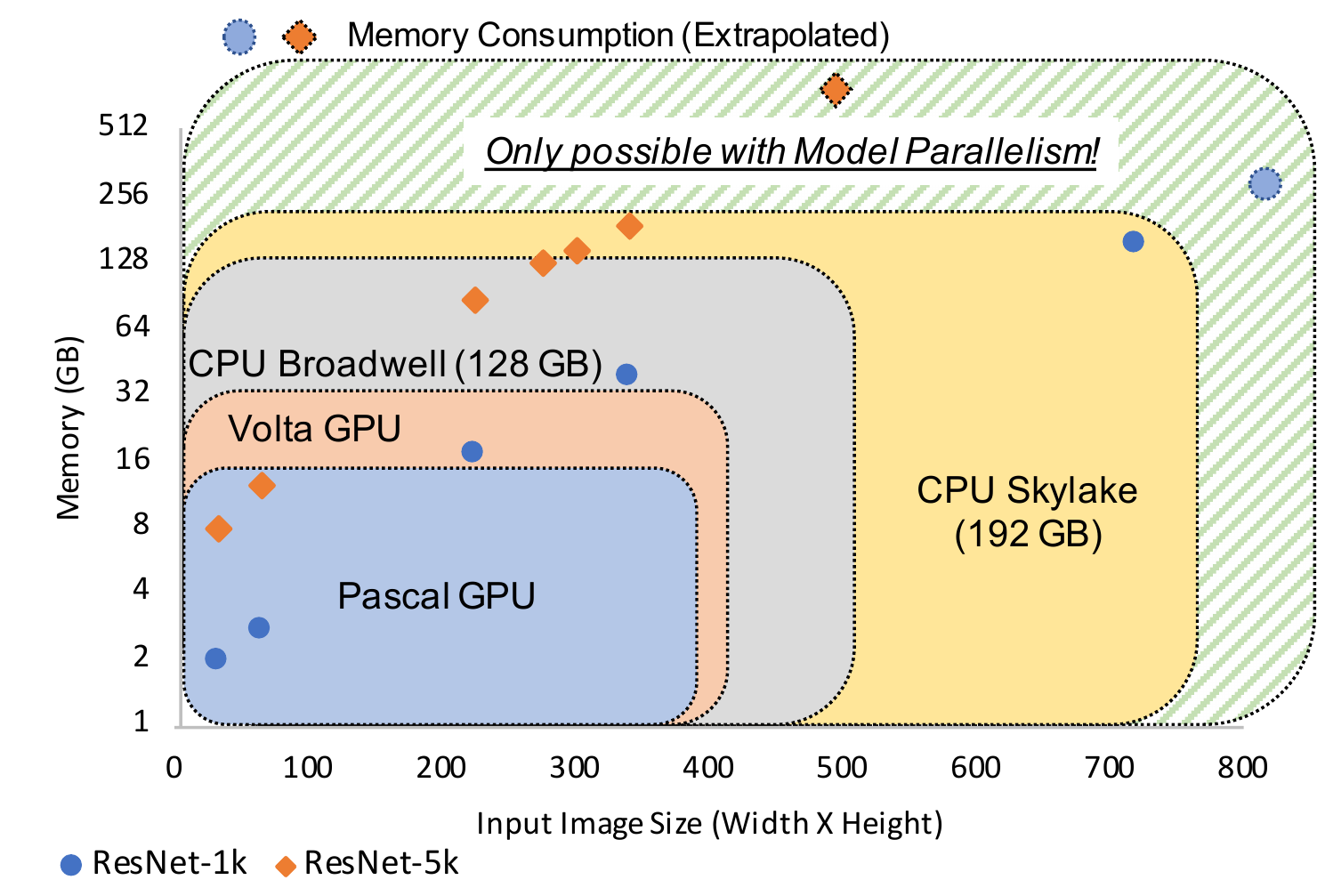}
    \vspace{-1.5ex}
    \caption{The Need for Model/Hybrid-Parallelism}
    \label{fig:need-for-mp}
    \vspace{-2.5ex}
\end{figure}

While data-parallel training offers good performance for models that can completely reside in the memory of a CPU/GPU, it \textit{can not} be used for models larger than the memory available. Larger and deeper models are being built to increase the accuracy of models even further~\cite{kerasio,resnet1k}. Figure~\ref{fig:need-for-mp} highlights how \textit{memory consumption} due to larger images and DNN depth limits the compute platforms that can be used for training; e.g. ResNet-1k~\cite{resnet1k} with the smallest possible batch-size of one (a single 224$\times$224 image) needs 16.8 GB memory and thus \textit{cannot} be trained on a 16 GB Pascal GPU. Similarly, ResNet-1k on image size 720$\times$720 needs 153 GB of memory, which makes it out-of-core for most platforms except CPU systems that have 192 GB memory. These \textit{out-of-core} models have triggered the \textit{need for model/hybrid parallelism}.

However, realizing \textit{model-parallelism}\footnote{Model-parallelism and layer-parallelism are equivalent terms when the smallest partition of a model is a layer~\cite{weird-trick,ben-nun}}---splitting the model (DNN) into multiple partitions --- is non-trivial and requires the knowledge of best practices in ML/DL as well as expertise in High Performance Computing (HPC). Little exists in the literature about model-parallelism for state-of-the-art DNNs like ResNet(s) on HPC systems. Combining data and model parallelism, also called hybrid-parallelism has received even less attention. Realizing model-parallelism and hybrid-parallelism efficiently is challenging because of \textit{four major problems}: 1) defining a distributed model is necessary but difficult because it requires knowledge of the model as well as of the underlying communication library and the distributed hardware, 2) implementing distributed forward/back-propagation is needed because partitions of the model now reside in different memory spaces and will need explicit communication, 3) obtaining parallel speedup on an inherently sequential task; forward pass followed by a backward pass, and 4) achieving scalability without losing out on a model's accuracy.

\begin{figure*}[htbp]
\centering
    \vspace{-2.0ex}
    \includegraphics[width=\textwidth]{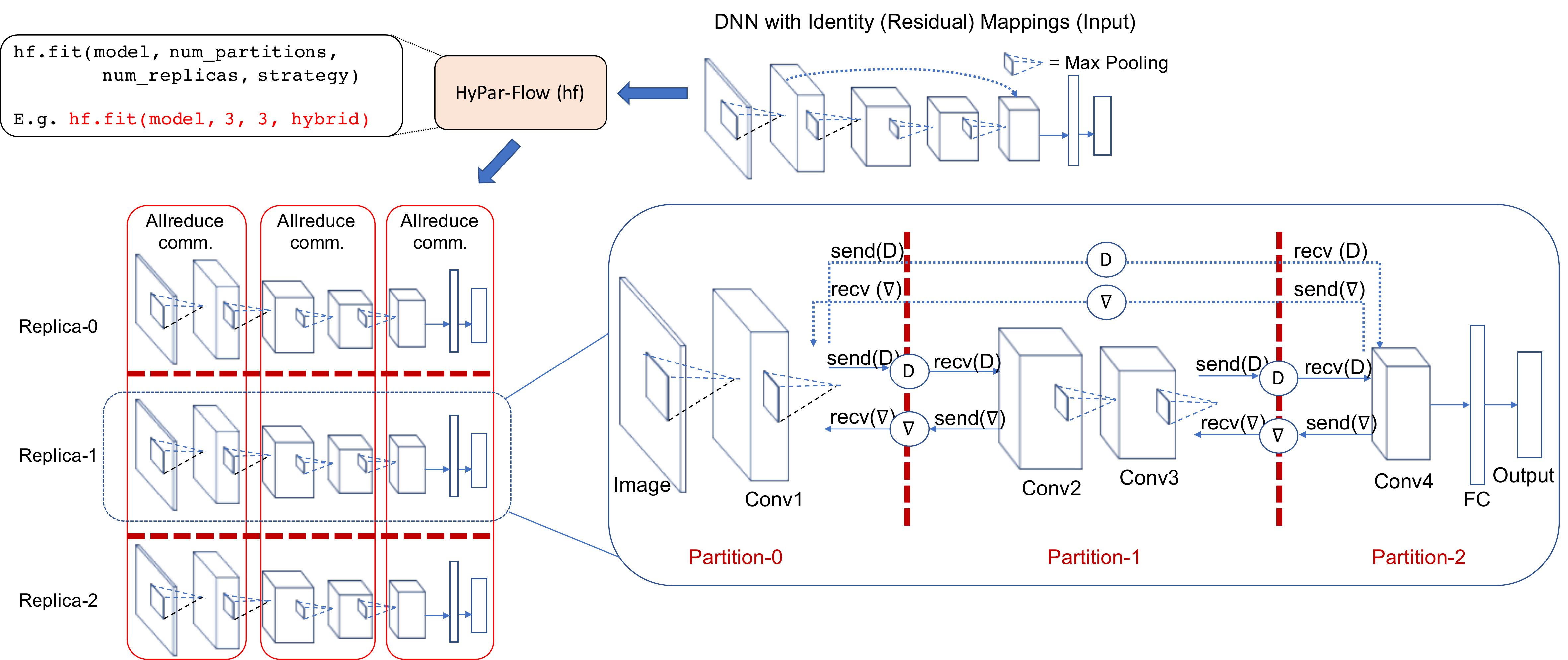}
    \caption{Proposed User-transparent Hybrid-Parallel Training Approach (HyPar-Flow)}
    \label{fig:teaser}
    \vspace{-2.5ex}
\end{figure*}

\vspace{1.0ex}
\noindent \textbf{Proposed Approach:} To address these four problems, we propose HyPar-Flow: a scalable, practical, and user-transparent system for hybrid-parallel training on HPC systems. We offer a simple interface that does not require any model-definition changes and/or manual partitioning of the model. Users provide four inputs: 1) A model defined using the Keras API, 2) Number of model partitions, 3) Number of model replicas, and 4) Strategy (data, model, or hybrid). Unlike existing systems, we design and implement all the cumbersome tasks like splitting the model into partitions, replicating it across processes, pipelining over batch partitions, and realizing communication inside HyPar-Flow. This enables the users to focus on the science of the model instead of system-level problems like the creation of model partitions and replicas, placement of partitions and replicas on cores and nodes, and performing communication between them. HyPar-Flow's simplicity from a user's standpoint and its complexity (hidden from the user) from our implementation's standpoint is shown in Figure~\ref{fig:teaser}.

\vspace{-1.0ex}
\MySubsection{Contributions}
\label{sec:contrib}

From a research and novelty standpoint, our proposed solution is both model-size as well as model-type agnostic. It is also different compared to all existing systems because we focus on high-level and abstract APIs like Keras that are used in practice instead of low-level tensors and matrices that are extremely hard to define state-of-the-art models with hundreds of layers. HyPar-Flow's solution to communication is also novel because it is the first system to exploit standard Message Passing Interface (MPI) primitives for inter-partition and inter-replica communication instead of reinventing single-use libraries. To the best of our knowledge, there are very few studies that focus on hybrid-parallel training of large DNNs; especially using TensorFlow and Keras in a user-transparent manner for HPC environments where MPI is a dominant programming model. We make the following key contributions in this paper:

\vspace{-1.0ex}
\begin{itemize}
    \item Analyze various model-definition APIs and DL frameworks and highlight why \textit{Keras} APIs and custom-built training loops using TensorFlow Eager's \textit{GradientTape} are well suited for realizing user-transparent hybrid-parallelism.
    
    \item Propose, design, and implement HyPar-Flow to enable parallel training of any Keras model (with consecutive as well as non-consecutive layer connections~\cite{ben-nun}) on multiple processes under any parallelization $<$strategy$>$, i.e. data, model, and hybrid.

    \item Thoroughly verify the correctness of the HyPar-Flow framework by training the models to state-of-the-art published accuracy.
       
    \item Evaluate HyPar-Flow's performance using a variety of models including VGG-16, ResNet-110, and ResNet-1001 on three HPC systems
    
    \item Report up to 3.1$\times$ speedup over sequential training for ResNet-110 and up to 1.6$\times$ speedup over data-parallel training for ResNet-1001 on a single node.
    
    \item Report 110$\times$ speedup over single-node on 128 Stampede2 nodes and 481$\times$ speedup over single-node on 512 Frontera nodes for ResNet-1001.
  
\end{itemize}
\vspace{-2.0ex}
\MySection{The Design Space for Parallel Training Frameworks}
\label{sec:related}

Alex Krizhevsky introduced model-parallelism on GPUs in~\cite{weird-trick} using a single-tower design that used data-parallelism in convolutional layers but model-paralle-lism in fully-connected layers. Simulation-based results about various parallelization strategies are presented in~\cite{amir-golami-spaa18}. The LBANN team presented model-parallel solutions including support for spatial convolutions split across nodes in~\cite{lbann-ipdps19}. However, model-parallelism in LBANN is not yet publicly available so we cannot compare its performance with HyPar-Flow. GPipe~\cite{gpipe} enables the training of extremely large models like AmoebaNet~\cite{amoebanet} on Google TPUs and accelerators. GPipe is publicly available but we found no examples and/or documentation to train models like ResNet(s) with model-parallel support on an HPC system. FlexFlow~\cite{flexflow} searches parallelization strategies using simulation algorithms and highlights different dimensions of parallelism in DNNs. FlexFlow uses Legion~\cite{legion} for communication within the node and GASNet across nodes. Unfortunately, FlexFlow only works on GPUs so we cannot offer a direct comparison. Also, we were unable to configure FlexFlow for multiple nodes. Mesh-TensorFlow (MTF)~\cite{tf-mesh-paper} is a language for distributed DL with an emphasis on tensors distributed across a processor mesh. MTF only works with the older TF APIs (sessions, graphs, etc.). Furthermore, the level at which MTF distributes work is much lower compared to HyPar-Flow, i.e., tensors vs. layers. Users of MTF need to re-write their entire model to be compatible with MTF APIs. Unlike MTF, HyPar-Flow works on the existing models without requiring any code/model changes. We summarize these related studies on data, model, and hybrid-parallelism and their associated features in Table~\ref{tab:related}. Out-of-core methods like~\cite{awan-hipc18,dragon} take a different approach to deal with large models, which is not directly comparable to model/hybrid-parallelism.

\vspace{-3.0ex}
\begin{table*}[htbp]%[r]{7.0cm}
\centering
\resizebox{0.9\textwidth}{!}{%
\begin{tabular}{|c|c|c|c|c|c|c|}
\hline
\multirow{2}{*}{\begin{tabular}[c]{@{}c@{}}\\Existing and \\ Proposed\\ Studies  \end{tabular}} & \multicolumn{6}{c|}{Features and Supported Platforms} \\ \cline{2-7} 
 & \begin{tabular}[c]{@{}c@{}}User\\ Transparent\end{tabular} & \begin{tabular}[c]{@{}c@{}}Speedup \\ over \\ Data-Parallel\end{tabular} & \begin{tabular}[c]{@{}c@{}}Communication\\ Runtime/Library\end{tabular} & \begin{tabular}[c]{@{}c@{}}Publicly\\ Available \\ MP Support\end{tabular} & \begin{tabular}[c]{@{}c@{}}Compatible w/\\ Keras\end{tabular} & \begin{tabular}[c]{@{}c@{}}Compatible w/\\ TF Eager\end{tabular} \\ \hline
AlexNet~\cite{alexnet,weird-trick} &  \xmark & \cmark & CUDA & \xmark & \xmark & \xmark \\ \hline
MXNet-MP~\cite{mxnet-mp} &  \xmark & Unknown & MPI & \cmark & \cmark & \xmark \\ \hline
LBANN~\cite{lbann-ipdps19} &  \cmark & \cmark & MPI/Aluminum & \xmark & \xmark & \xmark \\ \hline
Mesh TensorFlow~\cite{tf-mesh-paper} & \xmark &  \cmark & MPI & \cmark & \xmark & \xmark \\ \hline
Gpipe~\cite{gpipe} & \xmark & \xmark & gRPC/TF & \cmark & \xmark & Unknown \\ \hline
PipeDream~\cite{pipedream}& \xmark & \cmark & ZeroMQ & Unknown & \xmark & \xmark \\ \hline
FlexFlow~\cite{flexflow} & \cmark & \cmark & Legion/GASNet & \cmark & \xmark & \xmark \\ \hline
\textbf{\begin{tabular}[c]{@{}c@{}}Proposed \\ (HyPar-Flow)\end{tabular}} & \cmark & \cmark & MPI & Planned & \cmark & \cmark \\ \hline
\end{tabular}%
}
\vspace{1.0ex}
\caption{Features offered by HyPar-Flow compared to Existing Frameworks} 
\vspace{-6.5ex}
\label{tab:related}
\end{table*}

\MySection{Background}

\vspace{0.5ex}
We provide the necessary background in this section.

\vspace{1.0ex}
\noindent \textbf{DNN Training:} A DNN consists of different types of \textit{layers} such as convolutions (\textit{conv}), fully-connected or dense (\textit{FC}), pooling, etc. DNNs are usually trained using a labeled dataset. A full pass over this dataset is called an \textit{epoch} of training. Training itself is an iterative process and each iteration happens in two broad phases: 1) Forward pass over all the layers and 2) Back-propagation of loss (or error) in the reverse order. The end goal of DNN training is to obtain a model that has good prediction capabilities (\textit{accuracy}). To reach the desired/target \textit{accuracy} in the fastest possible time, the training process itself needs to be efficient. In this context, the total training time is a product of two metrics: 1) the number of epochs required to reach the target accuracy and 2) the time required for one epoch of training. 

\vspace{1.0ex}
\noindent \textbf{Data-Parallelism:} In data-parallel training, the complete DNN is replicated across all processes. However, the training dataset is partitioned across the processes. Since the model replicas on each of the processes train on different partitions of data, the weights (or parameters) learned are different on each process and thus need to be synchronized among replicas. In most cases, this is done by averaging the \textit{gradients} from all processes. This synchronization is performed by using a collective communication primitive like allreduce or by using parameter servers. The synchronization of weights is done at the end of every batch. This is referred to as \textit{synchronous parallel} in this paper. 

\vspace{1.0ex}
\noindent \textbf{Model and Hybrid-Parallelism:}
Data-parallelism works for models that can fit completely inside the memory of a single GPU/CPU. But as model sizes have grown, model designers have pursued aggressive strategies to make them fit inside a GPU's memory, which is a precious resource even on the latest Volta GPU (32 GB). This problem is less pronounced for CPU-based training as the amount of CPU memory is significantly higher (192 GB) on the latest generation CPUs. Nevertheless, some models can not be trained without splitting the model into partitions; Hence, model-parallelism is a necessity, which also allows the designers to come up with new models without being restricted to any memory limits. The entire model is partitioned and each process is responsible only for part (e.g. a layer or some layers) of the DNN. Model-parallelism can be combined with data-parallelism as well, which we refer to as hybrid-parallelism.
\MySection{Challenges in Designing Model and Hybrid-Parallelism}
\label{sec:challenges}

We expand on \textit{four problems} discussed earlier in Section~\ref{sec:intro} and elaborate specific challenges that need to be addressed for designing a scalable and user-transparent system like HyPar-Flow.

\vspace{0.5ex}
\noindent \textit{\textbf{\normalsize Challenge-1: Model-Definition APIs and Framework-specific Features}}
\noindent To develop a practical system like HyPar-Flow, it is essential that we thoroughly investigate APIs and features of DL frameworks. In this context, the design analysis of execution models like Eager Execution vs. Graph (or Lazy) Execution is fundamental. Similarly, analysis of model definition APIs like TensorFlow Estimators compared to Keras is needed because these will influence the design choices for developing systems like HyPar-Flow. Furthermore, the granularity of interfaces needs to be explored. For instance, using tensors to define a model is very complex compared to using a high-level model API like Keras and ONNX that follow the layer abstraction. Finally, we need to investigate the performance behavior of these interfaces and frameworks. Specific to HyPar-Flow, the main requirement from an API's perspective is to investigate a mechanism that allows us to perform user-transparent model partitioning. Unlike other APIs, Keras seems to provide us this capability via the \textit{tf.keras.Model} interface. 

\vspace{0.5ex}
\noindent \textit{\textbf{\normalsize Challenge-2: Communication between Partitions and Replicas}}

\noindent Data-parallelism is easy to implement as no modification is required to the forward pass or the back-propagation of loss (error) in the backward pass. However, for model-parallelism, we need to investigate methods and framework-specific functionalities that enable us to implement the forward and backward pass in a distributed fashion. To realize these, explicit communication is needed between model partitions. For hybrid-parallelism, even deeper investigation is required because communication between model replicas and model partitions needs to be well-coordinated and possibly overlapped. In essence, we need to design a distributed system, which embeds communication primitives like \textit{send}, \textit{recv}, and \textit{allreduce} for exchanging partial error terms, gradients, and/or activations during the forward and backward passes. An additional challenge is to deal with newer DNNs like ResNet(s)~\cite{resnet1k} as they have evolved from a linear representation to a more complex graph with several types of skip connections (shortcuts) like identity connections, convolution connections, etc. For skip connections, maintaining dependencies for layers as well as for model-partitions is also required to ensure deadlock-free communication across processes.

\vspace{0.5ex}
\noindent \textit{\textbf{\normalsize Challenge-3: Applying HPC Techniques to Improve Performance}}

\noindent Even though model-parallelism and hybrid-parallelism look very promising, it is unclear if they can offer performance comparable to data-parallelism. To achieve performance, we need to investigate if applying widely-used and important HPC techniques like 1) efficient placement of processes on CPU cores, 2) pipelining via batch splitting, and 3) overlap of computation and communication can be exploited for improving performance of model-parallel and hybrid-parallel training. Naive model-parallelism will certainly suffer from under-utilization of resources due to stalls caused by the sequential nature of computation in the forward and backward passes.
\MySection{HyPar-Flow: Proposed Architecture and Designs}
\label{sec:design}

We propose HyPar-Flow as an abstraction between the high-level ML/DL frameworks like TensorFlow and low-level communication runtimes like MPI as shown in Figure~\ref{fig:overview}. The HyPar-Flow middleware is directly usable by ML/DL applications and no changes are needed to the code or the DL framework. The four major internal components of HyPar-Flow, shown in Figure~\ref{fig:design-details}, are 1) Model Generator, 2) Trainer, 3) Communication Engine (CE), and 4) Load Balancer. 

\vspace{-5.5ex}
\mytwocolfigcustom  
{Overview of the Execution Stack}
{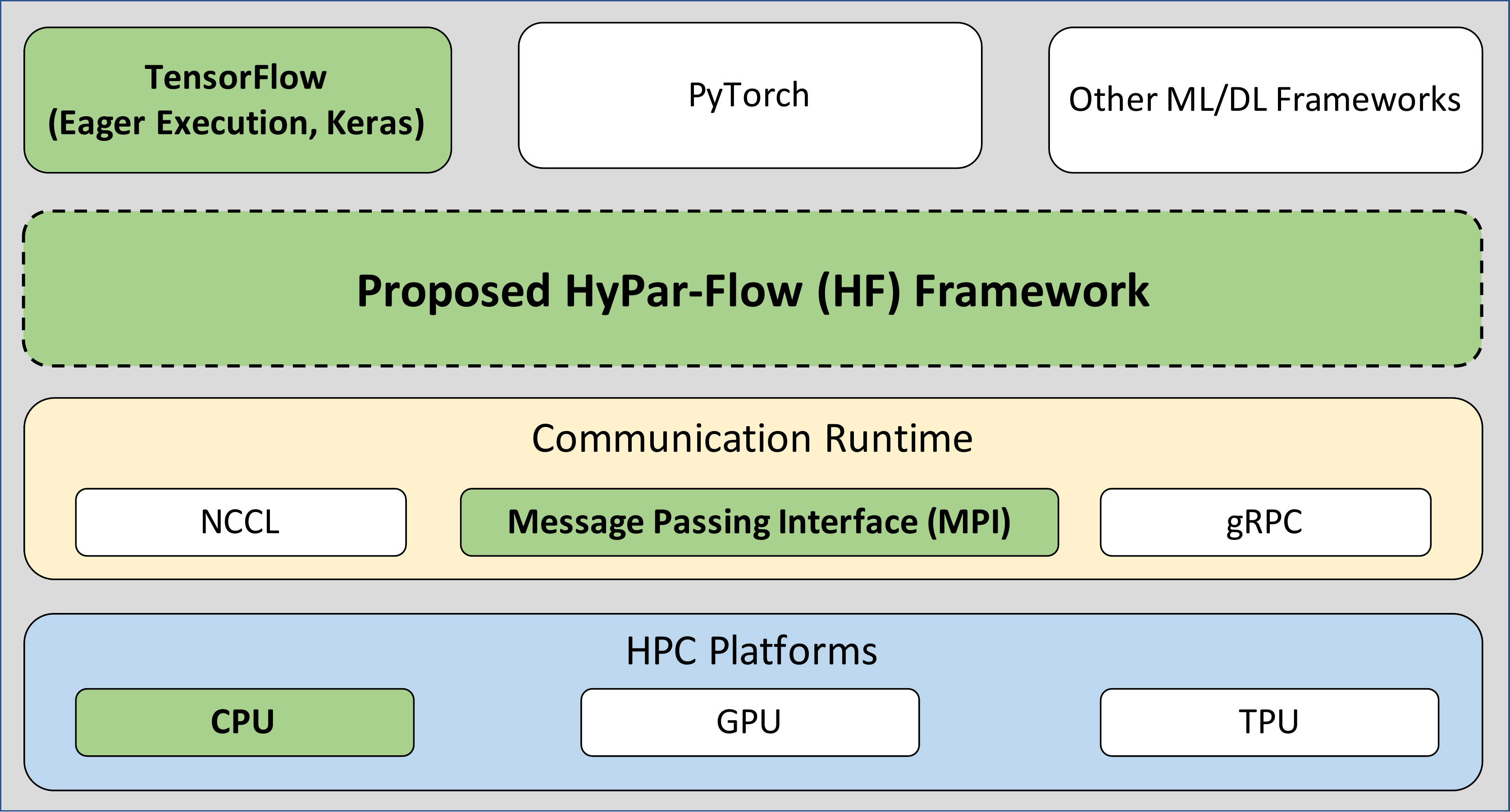}
{fig:overview}
{Major Components of HyPar-Flow}
{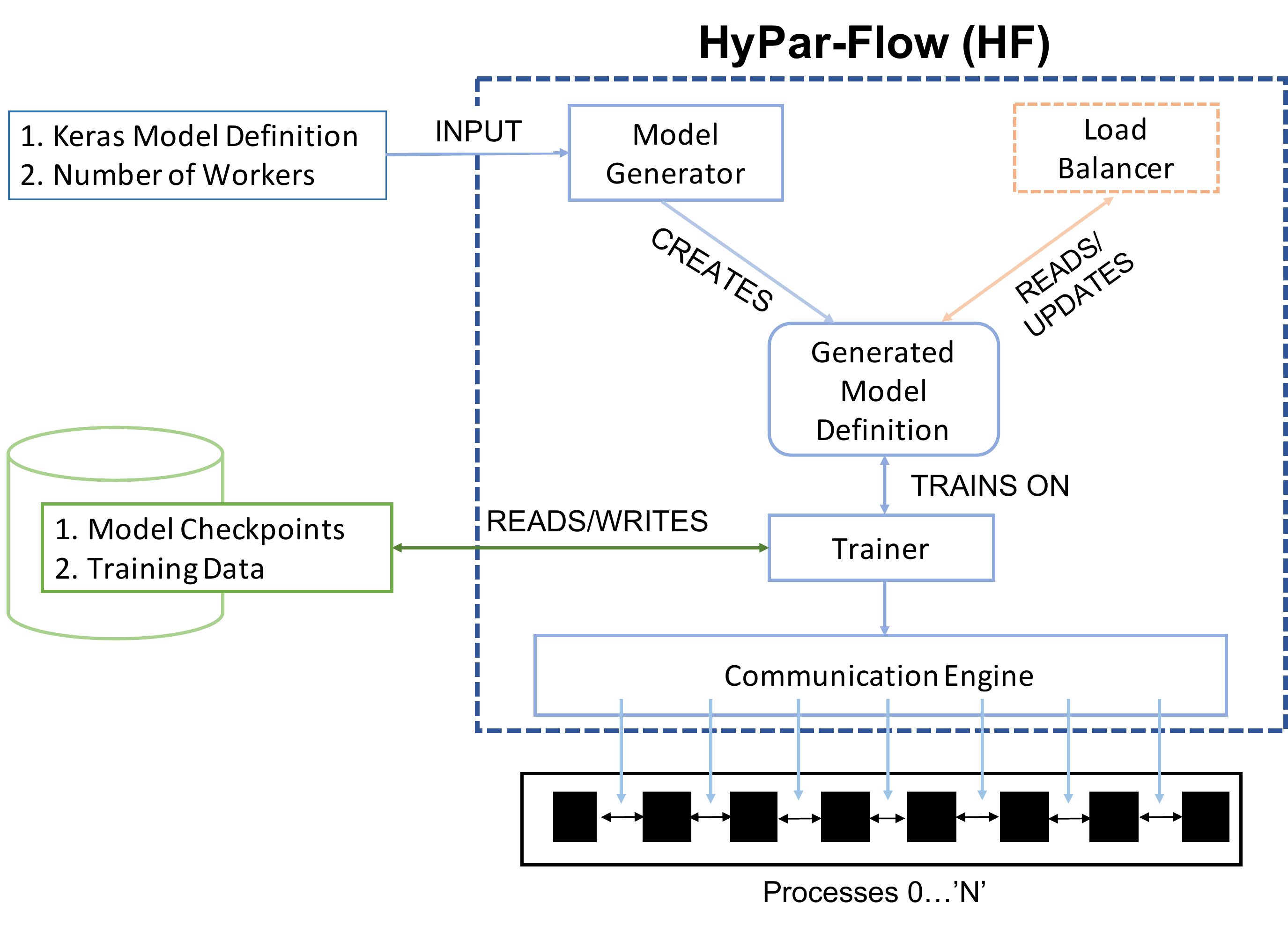}
{fig:design-details}
{HyPar-Flow: A Middleware for Hybrid-Parallel Training}
\vspace{-4.0ex}

\textit{The subsections that follow provide details of design schemes and strategies for HyPar-Flow and challenges (\textbf{C1--C3}) addressed by each scheme.}

\vspace{-0.5ex}
\MySubsection{Designing Distributed Model Representation (Address C1)} 
\label{sec:modelgen}

The \textit{Model Generator} component is responsible for creating an internal representation of a DNN (e.g. a Keras model) suitable for distributed training (Figure~\ref{fig:teaser}). In the standard single-process (sequential) case, all trainable variables (or weights) of a model exist in the address space of a single process so calling \textit{tape.gradients()} on a \textit{tf.GradientTape} object to get gradients will suffice. However, this is not possible for model-parallel training as trainable variables (weights) are distributed among model-partitions. To deal with this, we first create a local model object on all processes using the \textit{tf.keras.model} API. Next, we identify the layers in the model object that are local to the process. Finally, we create dependency lists that allow us to maintain layer and rank dependencies for each of the local model's layers. These three components define our internal distributed representation of the model. This information is vital for realizing distributed back-propagation (discussed next) as well as for other HyPar-Flow components like the \textit{Trainer} and the \textit{Communication Engine}. 

\MySubsection{Implementing Distributed Back-Propagation (Address C1,C2)}
\label{sec:trainer}

Having a distributed model representation is crucial. However, it is only the first step. The biggest challenge for HyPar-Flow and its likes are: ``How to train a model that is distributed across process boundaries?''. We deal with this challenge inside the \textit{Trainer} component. First, we analyze how training is performed on a standard (non-distributed) Keras model. Broadly, there are two ways to do so: 1) \textit{model.fit(..)} and 2) \textit{model.train\_on\_batch(..)}. Second, we explore how we can design an API that is very similar to the standard case. To this end, we expose a single \textit{hf.fit(..)} interface that takes parallelization strategy as an argument. The value of the \textit{strategy} argument can be model, data, or hybrid. Third, we design a custom training loop for distributed back-propagation for the model/hybrid parallel case. For data-parallel, it is not needed because the model is replicated on all processes instead of being distributed across processes.

\begin{figure}[htbp]
    \centering
    \vspace{-5.0ex}
    \includegraphics[width=0.6\columnwidth]{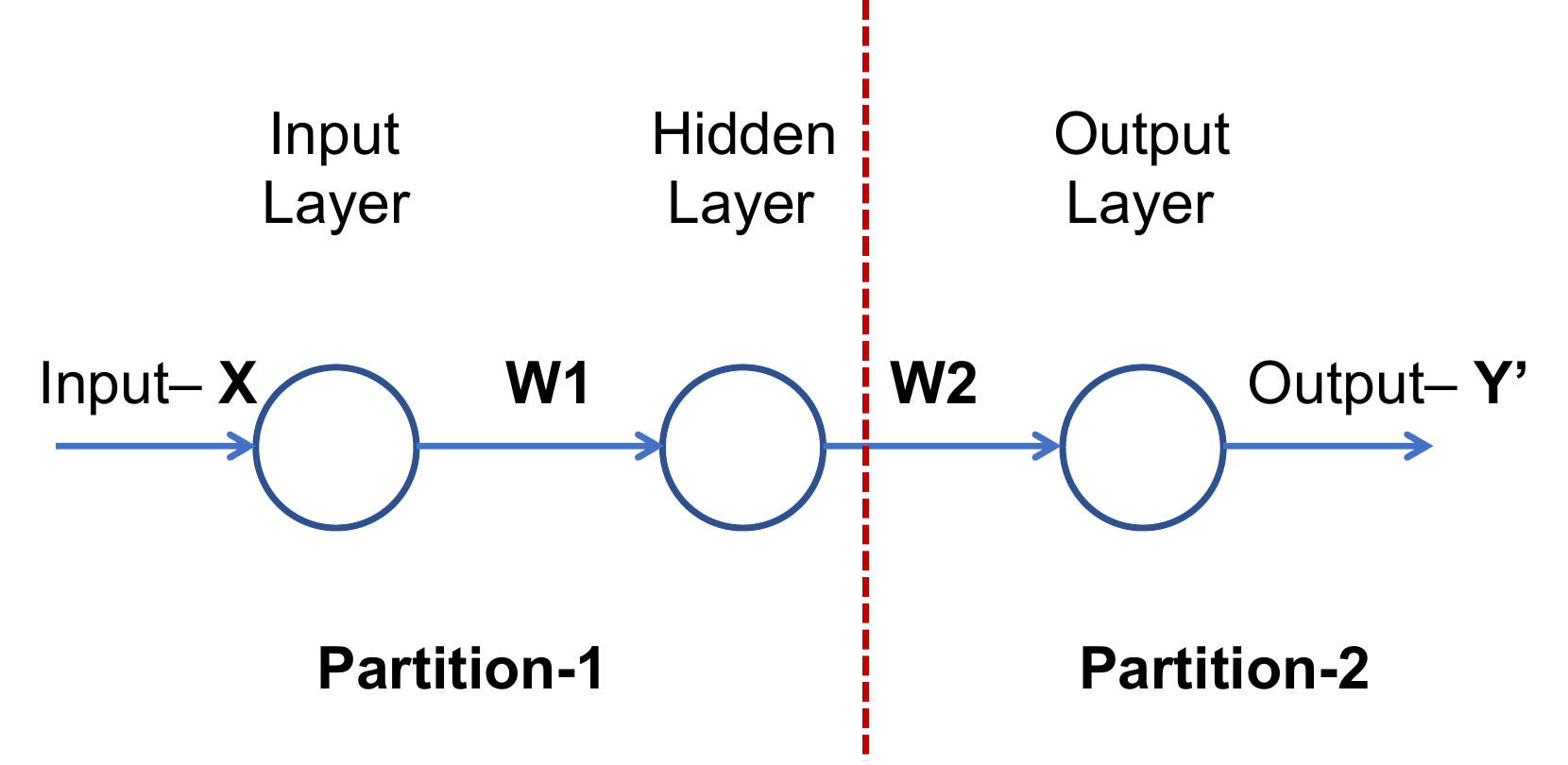}
    \vspace{-2.0ex}
    \caption{A Neural Network with a single Hidden layer}
    \vspace{-2.0ex}
    \label{fig:simplenet}
\end{figure}

We show a very simple DNN in Figure~\ref{fig:simplenet} to explain back-propagation and highlight what needs to be done for realizing a distributed version. In addition to Figure~\ref{fig:simplenet}, we use Equations 1--7 to provide a more detailed explanation. There are three key data elements in DNN Training: 1) The input $X$, 2) The predicted output $Y'$, and 3) The actual output (or label) $Y$. The intermediate output from the hidden layer is denoted as $V$. The difference between $Y$ and $Y'$ is called error or loss labeled as $L$ (Eq.~\ref{eq:2}).

\hrule
\hrule
%0
\begin{equation}
Y = ActualOutput, Y' = PredictedOutput
\label{eq:1}
\end{equation}
\vspace{-1.5ex}
%1
\begin{equation}
L (Loss) =loss\_function(Y, Y')
\label{eq:2}
\end{equation}
\vspace{-1.5ex}
%2
\begin{equation}
V (Hidden Layer)=W_{1} (Weight-on-hidden-layer) *X (Input)
\label{eq:3}
\end{equation}
\vspace{-1.5ex}
%3
\begin{equation}
Y' (PredictedOutput)=W_{2} (Weight-on-output-layer)*V
\label{eq:4}
\end{equation}
%\vspace{-1.5ex}
%4
\begin{equation}
D2= \frac{\partial L}{\partial W_{2}}=\frac{\partial L}{\partial Y'}*\frac{\partial Y'}{\partial W_{2}}
\label{eq:5}
\end{equation}
\vspace{-1.5ex}
%5
\begin{equation}
D1= \frac{\partial L}{\partial W_{1}}= partial\_error *\frac{\partial V}{\partial W_{1}}
\label{eq:6}
\end{equation}
%6
\begin{equation}
    \label{eq:7}
    partial\_error=\frac{\partial L}{\partial Y'}*\frac{\partial Y'}{\partial V}
\end{equation}
\hrule
\hrule
\vspace{1.5ex}

To realize distributed back-propagation, we need 1) partial derivative (D1) of Loss $L$ with respect to the weight $W1$, and 2) partial derivative (D2) of Loss $L$ with respect to the weight $W2$. The challenge for multi-process case is that the term called ``partial error'' shown in Equations~\ref{eq:6} and~\ref{eq:7} can only be calculated on \textit{Partition-2} (Figure~\ref{fig:simplenet}) as $Y'$ only exists there. To calculate D1, \textit{Partition-1} needs this ``partial error'' term in addition to D1. Because we rely on accessing gradients using the DL framework's implementation, this scenario poses a fundamental problem. TensorFlow, the candidate framework for this work, does not provide a way to calculate gradients that are not part of a layer. To implement this functionality, we introduce the notion of \textit{grad layer} in HyPar-Flow, which acts as a pseudo-layer inserted before the actual layer on each model-partition. We note that TensorFlow's \textit{GradientTape} cannot be directly used for this case. \textit{Grad layers} ensure that we can call \textit{tape.gradients()} on this \textit{grad layer} to calculate the partial errors during back-propagation. Specifically, a \textit{grad layer} is required for each \textit{recv} operation so that partial error can be calculated for each preceding partition's input. A call to tape.gradients() will return a list that contains gradients as well as partial errors. The list is then used to update the model by calling \textit{optimizer.apply\_gradients()}.

We note that there is no need to implement distributed back-propagation for the data-parallel case as each model-replica is independently performing the Forward and Backward pass. The gradients are only synchronized (averaged) at the end of the Backward pass (back-propagation) using \textit{allreduce} to update the model weights in a single step. 

\MySubsection{Realizing Inter-Partition/-Replica Comm. (Address C2,C3)}
\label{sec:ce}

In Sections~\ref{sec:modelgen} and~\ref{sec:trainer}, we discussed how the distributed model definition is generated and how back-propagation can be implemented for a model that is distributed across processes. However, \textit{Trainer} and \textit{Model Generator} only provide an infrastructure for distributed training. The actual communication of various types of data is realized in HyPar-Flow's \textit{Communication Engine (CE)}. The CE is a light-weight abstraction for internal usage and it provides four simple APIs: 1) send, 2) recv, 3) broadcast and 4) allreduce. 

\vspace{1.0ex}
\noindent \textbf{HyPar-Flow CE Basic Design:} For pure data-parallelism, we only need to use allreduce. However, for model-parallelism, we also need to use point-to-point communication between model-partitions. In the forward pass, the send/recv combination is used to propagate \textit{partial predictions} from each partition to the next partition starting at \textit{Layer 1}. On the other hand, send/recv is used to back-propagate the \textit{loss} and \textit{partial-errors} from one partition to the other starting at \textit{Layer N}. Finally, for hybrid-parallelism, we need to introduce allreduce to accumulate (average) the gradients across model replicas. We note that this is different from the usage of allreduce in pure data-parallelism because in this case, the model itself is distributed across different partitions so allreduce cannot be called directly on all processes. One option is to perform another p2p communication between model replicas for gradient exchange. The other option is to exploit the concept of MPI communicators. We choose the latter one because of its simplicity as well as the fact the MPI vendors have spent considerable efforts to optimize the allreduce collective for a long time. To realize this, we consider the same model-partition for all model-replicas to form the \textit{Allreduce communicator}. Because we only need to accumulate the gradients local to a partition across all replicas, allreduce called on this communicator will suffice. Please refer back to Figure~\ref{fig:teaser} (Section~\ref{sec:intro}) for a graphical illustration of this scheme.

\vspace{0.5ex}
\noindent \textbf{HyPar-Flow CE Advanced Design:} The basic CE design described above works but does not offer good performance. To push the envelope of performance further, we investigate two HPC optimizations: 1) we explore if the overlap of computation and communication can be exploited for all three parallelization strategies and 2) we investigate if pipelining can help overcome some of the limitations that arise due to the sequential nature of the forward/backward passes. Finally, we also handle some advanced cases for models with non-consecutive layer connections (e.g. ResNet(s)), which can lead to deadlocks. 

\vspace{0.5ex}
\noindent \textbf{\textit{Exploiting Overlap of Computation and Communication:}} To achieve near-linear speedups for data-parallelism, the overlap of computation (forward/ backward) and communication (allreduce) has proven to be an excellent choice. Horovod, a popular data-parallelism middleware, provides this support so we simply use it inside HyPar-Flow for pure data-parallelism. However, for hybrid-parallelism, we design a different scheme. We create one MPI communicator per model partition whereas the size of each communicator will be equal to the number of model-replicas. This design allows us to overlap the allreduce operation with the computation of other partitions on the same node. An example scenario clarifies this further: if we split the model across 48 partitions, then we will use 48 allreduce operations (one for each model-partition) to get optimal performance. This design allows us to overlap the allreduce operation with the computation of other partitions on the same node. 

\vspace{0.5ex}
\noindent \textbf{\textit{Exploiting Pipeline Stages within Each Minibatch:}} Because DNN training is inherently sequential, i.e., the computation of each layer is dependent on the completion of the previous layer. This is true for the forward pass, as well as for the backward pass. To overcome this performance limitation, we exploit a standard technique called pipelining. The observation is that DNN training is done on batches (or mini-batches) of data. This offers an opportunity for pipelining as a training step on samples within the batch is parallelizable. Theoretically, the number of pipeline stages can be varied from 1 all the way to batch size. This requires tuning or a heuristic and will vary according to the model and the underlying system. Based on hundreds of experiments we performed for HyPar-Flow, we derive a simple heuristic: use the largest possible number for pipeline stages and decrease it by a factor of two. In most cases, we observed that $num\_pipeline\_stages$ $=$ $batch\_size$ provides the best performance.

\begin{figure}[htbp]
\centering
    \vspace{-2.0ex}    
    \includegraphics[width=0.7\textwidth]{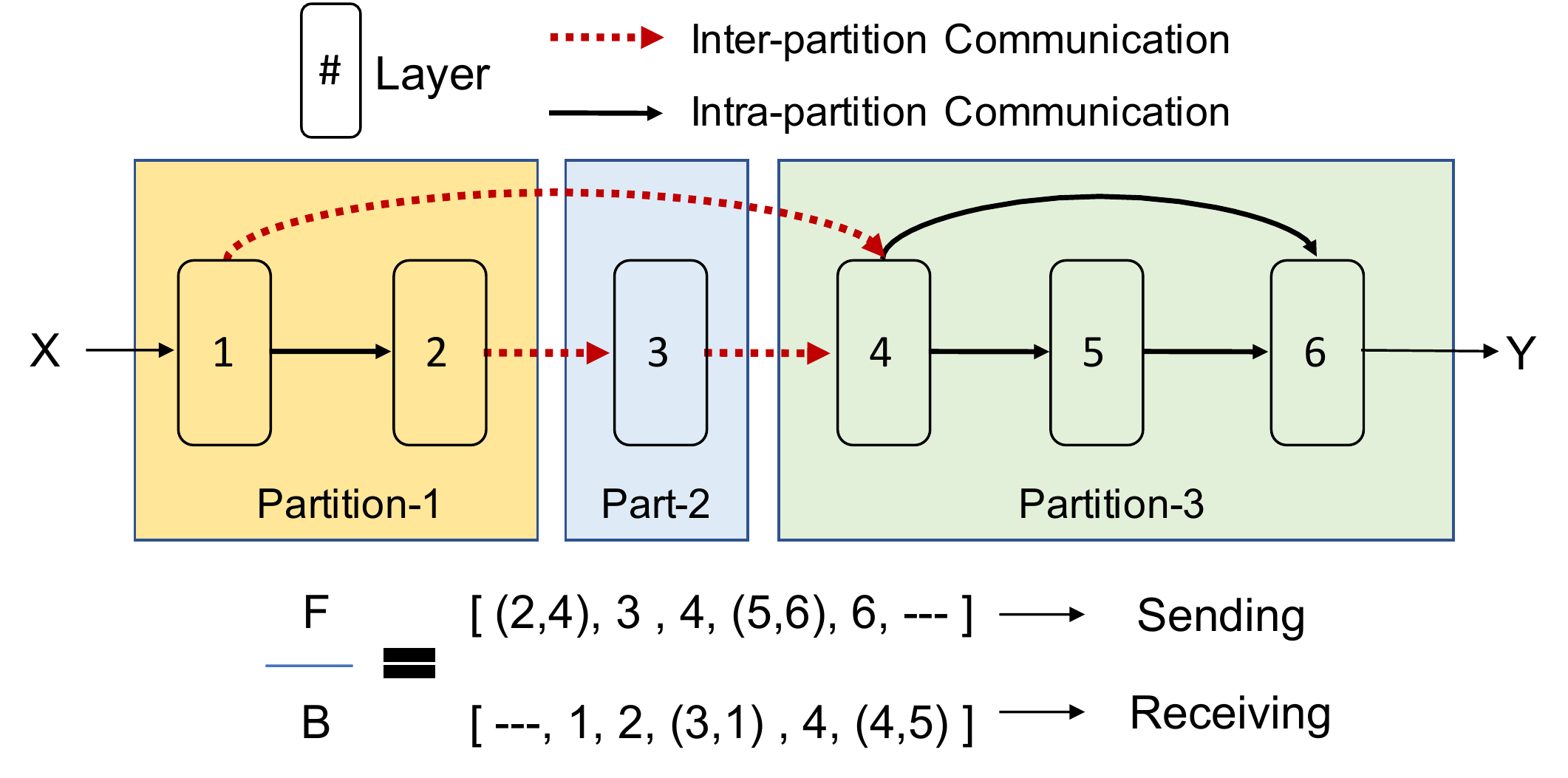}
    \vspace{-2.0ex}
    \caption{Avoiding Deadlocks for Models with Non-consecutive Connections}%The direction of arrows shown in the figure is to illustrate communication in the Forward Pass only. The communication will happen in the reverse direction for the Backward Pass.}
    \label{fig:deadlock-avoidance}
    \vspace{-5.0ex}
\end{figure}

\vspace{0.5ex}
\noindent \textbf{\textit{Special Handling for Non-consecutive Models:}} Figure~\ref{fig:deadlock-avoidance} shows a non-consecutive model with skip connections that requires communication 1) between adjacent model-partitions for boundary layers and 2) non-adjacent model-partitions for the skip connections. To handle communication dependencies among layers for each model-partition, we create two lists: 1) Forward list and 2) Backward list. Each list is a list of lists to store dependencies between layers as shown in Figure~\ref{fig:deadlock-avoidance}. ``F'' corresponds to the index of the layer to which the current layer is sending its data and ``B'' corresponds to the index of the layer from which the current layer is receiving data. An arbitrary sequence of sending and receiving messages may lead to a deadlock. For instance, if \textit{Partition-1} sends the partial predictions to \textit{Partition-3} when \textit{Partition-3} is waiting for predictions from \textit{Partition-2}, a deadlock will occur as \textit{Partition-2} is itself blocked (waiting for results from \textit{Partition-1}). To deal with this, we sort the message sequence according to the ranks so that the partition sends the first message to the partition which has the next layer. 

\MySubsection{Load Balancer}
The models we used did not show any major load imbalance but we plan to design this component in the future to address emerging models from other application areas that require load balancing capabilities from HyPar-Flow.
\MySection{Performance Characterization and Correctness Testing}

We have used three HPC systems to evaluate the performance and test the correctness of HyPar-Flow: 1) \textbf{Frontera} at Texas Advanced Computing Center (TACC), 2) \textbf{Stampede2} (Skylake partition) at TACC, and 3) \textbf{Epyc}: A local system with dual-socket AMD EPYC 7551 32-core processors. 

\vspace{0.5ex}
\noindent \textbf{Inter-connect:} Frontera nodes are connected using Mellanox InfiniBand HDR-100 HCAs whereas Stampede2 nodes are connected using Intel Omni-Path HFIs.

\vspace{0.5ex}
\noindent \textbf{DL Framework:} All experiments have been performed using TensorFlow v1.13.

\vspace{0.5ex}
\noindent \textbf{MPI Library:} MVAPICH2 2.3.2 was used on Frontera, Intel MPI 2018 was used on Stampede2, and MVAPICH2 2.3.1 was used on Epyc.

\vspace{0.5ex}
\noindent \textbf{Model Definitions:} We use and modify model definitions for VGG and ResNet(s) presented in Keras Applications/Examples~\cite{kerasio}.

\vspace{0.5ex}
\noindent \textbf{Note about GPUs:} The design schemes proposed for HyPar-Flow are architecture-agnostic and can work on CPUs and/or GPUs. However, in this paper, we focus only on designs and scale-up/scale-out performance of many-core CPU clusters. We plan to perform in-depth GPU-based HyPar-Flow studies in the future. 

\vspace{1.2ex}
\textit{We now present correctness related experiments followed by a comprehensive performance evaluation section.}

\MySubsection{Verifying the Correctness of HyPar-Flow}
\label{sec:verif}

Because we proposed and designed HyPar-Flow as a new system built from scratch, it is important to provide confidence to the users that HyPar-Flow not only offers excellent performance but also correctly trains the model. To this end, we present the correctness results based on two types of accuracy-related metrics: 1) {Train Accuracy (train\_acc)- Percentage of correct predictions for the training data during the training process and 2) Test Accuracy (test\_acc)- Percentage of correct predictions for the testing data on the trained model. Both metrics are covered for small scale training using \textbf{VGG-16} on the CIFAR-10 dataset. We train VGG-16 for 10 epochs using 8 model-partitions on two Stampede2 nodes with a batch size of 128 and 16 pipeline stages as shown in Figure~\ref{fig:vgg-parts-acc}. Next, we show test accuracy for \textbf{ResNet-110-v1} in Figure~\ref{fig:res110-parts-acc} and \textbf{ResNet-1001-v2} in Figure~\ref{fig:resnet-acc-compare}. The learning rate (LR) schedule was used from Keras Applications~\cite{kerasio} for both ResNet(s) and was kept similar for sequential as well as parallel training variants. Training for ResNet-110 and ResNet-1001 was performed for 150  and 50 epochs, respectively. The following variants have been compared:

\noindent 1) \textbf{SEQ (GT)} - Sequential using tf.GradientTape (GT).

\noindent 2) \textbf{SEQ (MF)} - Sequential using model.fit (MF). 

\noindent 3) \textbf{SEQ (MF-E)} - Sequential using model.fit (MF) and (E)ager Execution.

\noindent 4) \textbf{HF-MP (2)/(56)} - HyPar-Flow model-parallel with 2/56 model-partitions.

\mythreecolfig
{VGG-16 Training (all metrics) for 10 epochs with BS=128 and LR=0.0002}
{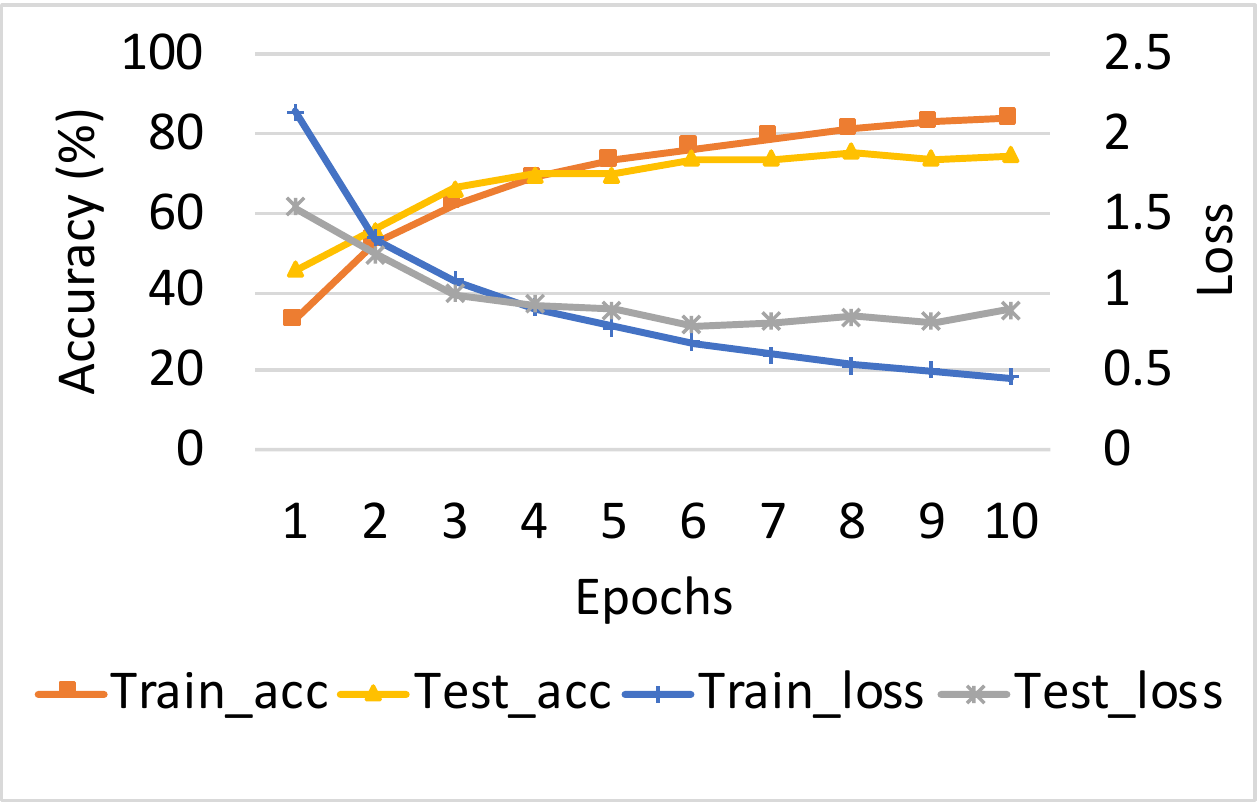}
{fig:vgg-parts-acc}
{ResNet-110-v1 Test Accuracy for 150 Epochs with BS=32}
{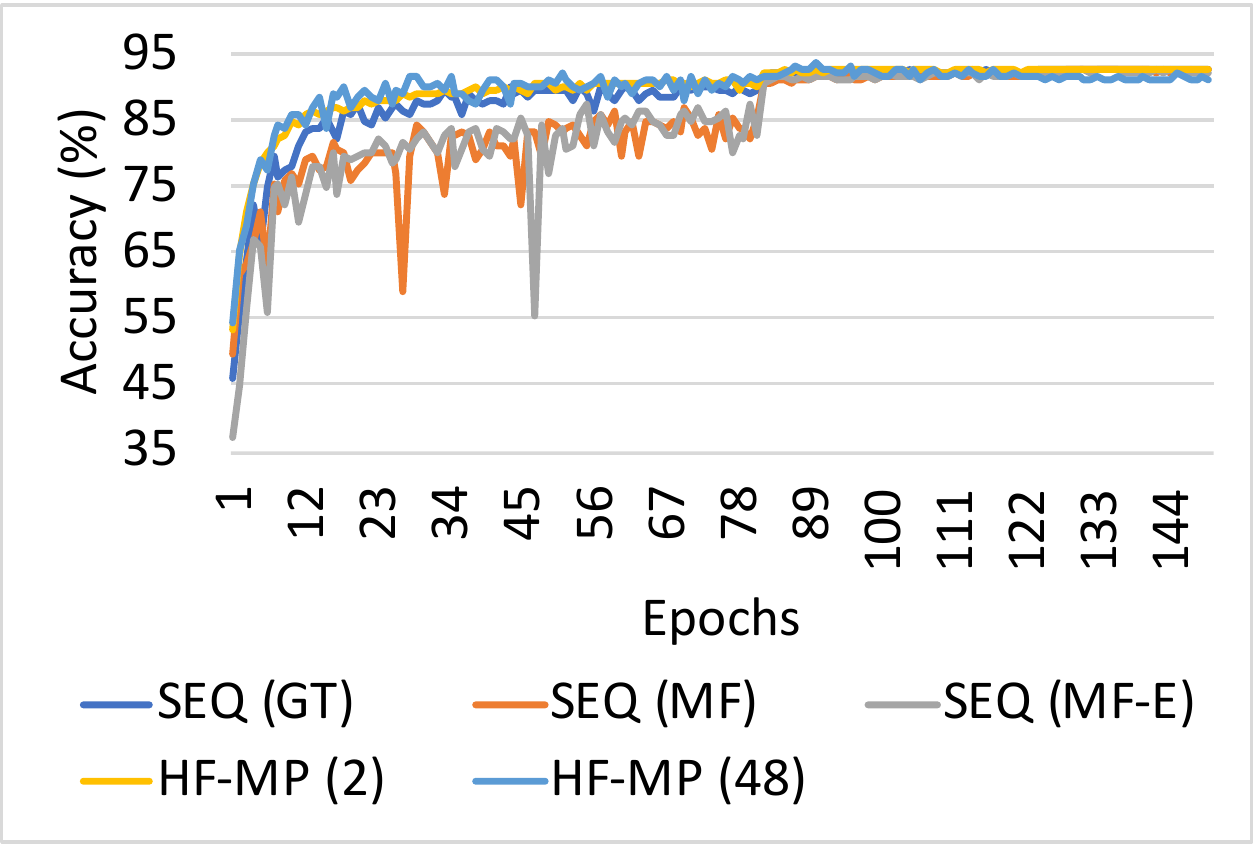}
{fig:res110-parts-acc}
{ResNet-1001-v2 Test Accuracy for 50 epochs with BS=32}
{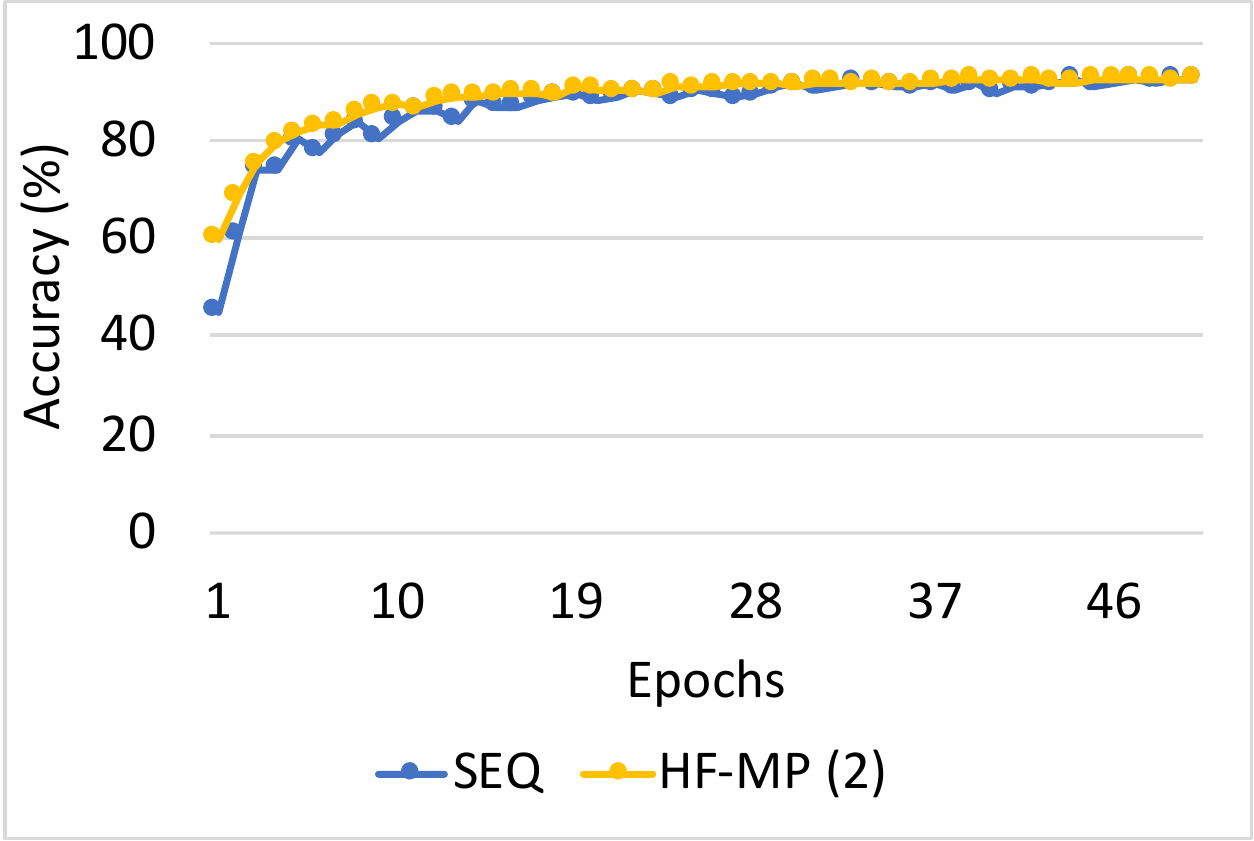}
{fig:resnet-acc-compare}
%{Verifying Correctness of HyPar-Flow}

\vspace{1.5ex}
\noindent \textbf{Discussion:} Clearly, model-parallel training with HyPar-Flow is meeting the accuracy of the sequential model for 150 and 50 epochs of training for ResNet-110 and ResNet-1001, respectively. We note that training is a stochastic process and there are variations in earlier epochs whether we use the sequential version or the model-parallel version. However, the significance is of the end result, which in this case peaks at 92.5\% for all the configurations presented. We ran multiple training jobs to ensure that the trends presented are reproducible.

\MySubsection{Experimental Setup for Performance Evaluation} 

We use the term ``process'' to refer to a single \textit{MPI Process} in this section. The actual mapping of the process to the compute units (or cores) varies according to the parallelization strategy being used. \textit{Images/second} (or Img/sec) is the \textit{metric} we are using for performance evaluation of different types of training experiments. Number of images processed by the DNN during training is affected by the \textit{depth} (number of layers) of the model, batch size (\textit{bs}), image size (W$\times$H), and number of processes. Higher \textit{Img/sec} indicates better performance. Some important terms are clarified further:

\vspace{0.5ex}
\noindent \textbf{Batch Size (BS):} \# of samples in the batch (mini-batch)

\noindent \textbf{Effective Batch Size (EBS)} = BS $\times$ num\_replicas for data/hybrid parallelism 

\noindent \textbf{Effective Batch Size (EBS)} = BS for model-parallelism 

\noindent\textbf{Image Size:} Dimension of the image (Width$\times$Height).

\vspace{0.5ex}
\noindent \textbf{Legend Entries for Graphs} in Sections~\ref{sec:single} and ~\ref{sec:multi-1} are:
\vspace{-1.5ex}
\begin{itemize}
    \item \textbf{Sequential}: Single-process DNN training using default TF/Keras APIs.
    \item \textbf{HF (MP)}: DNN training using \textit{hf.fit (..,strategy=model-parallel)}
    \item \textbf{HF (DP)}: DNN training using \textit{hf.fit(..,strategy=data-parallel)}
    \item \textbf{Horovod (DP)}: DNN training using Horovod directly (data-parallel)
\end{itemize}
\vspace{-1.5ex}

\MySubsection{Model-Parallelism on a Single Node}
\label{sec:single}

We train various models on a single Stampede2 node-- dual-socket Xeon Skylake with 48 cores and 96 threads (hyper-threading enabled). The default version of TensorFlow relies on underlying math libraries like OpenBLAS and Intel MKL. On Intel systems, we tried the Intel-optimized version of TensorFlow, but it failed with different errors such as ``function not implemented'' etc. For the AMD system, we used the OpenBLAS available on the system. Both of these platforms offer very slow sequential training. \textit{We present single-node results for VGG-16, ResNet-110-v1, and ResNet-1001-v2}. 

\vspace{1.0ex}
\noindent \textbf{VGG-16} has 16 layers so it can be split in to as many as 16 partitions. We try all possible cases and observe the best performance for num\_partitions=8. As shown in Figure~\ref{fig:vgg-single}, we see that HF (MP) offers better performance for small batch sizes and HF/Horovod (DP) offers better performance for large batch sizes. HF (MP) offers better performance compared to sequential (1.65$\times$ better at BS 1024) as well as to data-parallel training (1.25$\times$ better at BS 64) for VGG-16 on Stampede2.

\noindent \textbf{ResNet-110-v1} has 110 layers so we were able to exploit up to 48 model-partitions within the node as shown in Figure~\ref{fig:res110-single}. We observe the following: 1) HF (MP) is up to 2.1$\times$ better than sequential at BS=1024, 2) HF (MP) is up to 1.6 $\times$ better than Horovod (DP) and HF (DP) at BS=128, and 3) HF (MP) is 15\% slower than HF (DP) at BS=1024. The results highlight that model-parallelism is better at smaller batch sizes and data-parallelism are better only when large batch-size is used. Figure~\ref{fig:epyc} shows that HF (MP) can offer up to 3.2$\times$ better performance than sequential training for ResNet-110-v1 on \textit{Epyc} (64 cores). \textit{Epyc} offered better scalability with increasing batch sizes compared to Stampede2 nodes (Figure~\ref{fig:res110-single} vs.~\ref{fig:epyc}) The performance gains suggest that HF (MP) can better utilize all cores on \textit{Eypc} compared to sequential training.

\vspace{1.0ex}
\noindent \textbf{ResNet-1001-v2:} To push the envelope of model depth and stress the proposed HyPar-Flow system, we also perform experiments for ResNet-1001-v2, which has 1,0001 layers and approximately 30 million parameters. Figure~\ref{fig:res1k-single} shows the performance for ResNet-1001-v2. It is interesting to note that data-parallel training performs poorly for this model. This is because the number of parameters increases the synchronization overhead for HF (DP) and Horovod (DP) significantly. Hence, even for large batch sizes, the computation is not enough to amortize the communication overhead. Thus, HF (MP) offers much better performance compared to sequential (2.4$\times$ better at BS = 256) as well as to data-parallel training (1.75$\times$ better at BS = 128).

\mytwocolfig
{VGG-16 up to 8 model-partitions}
{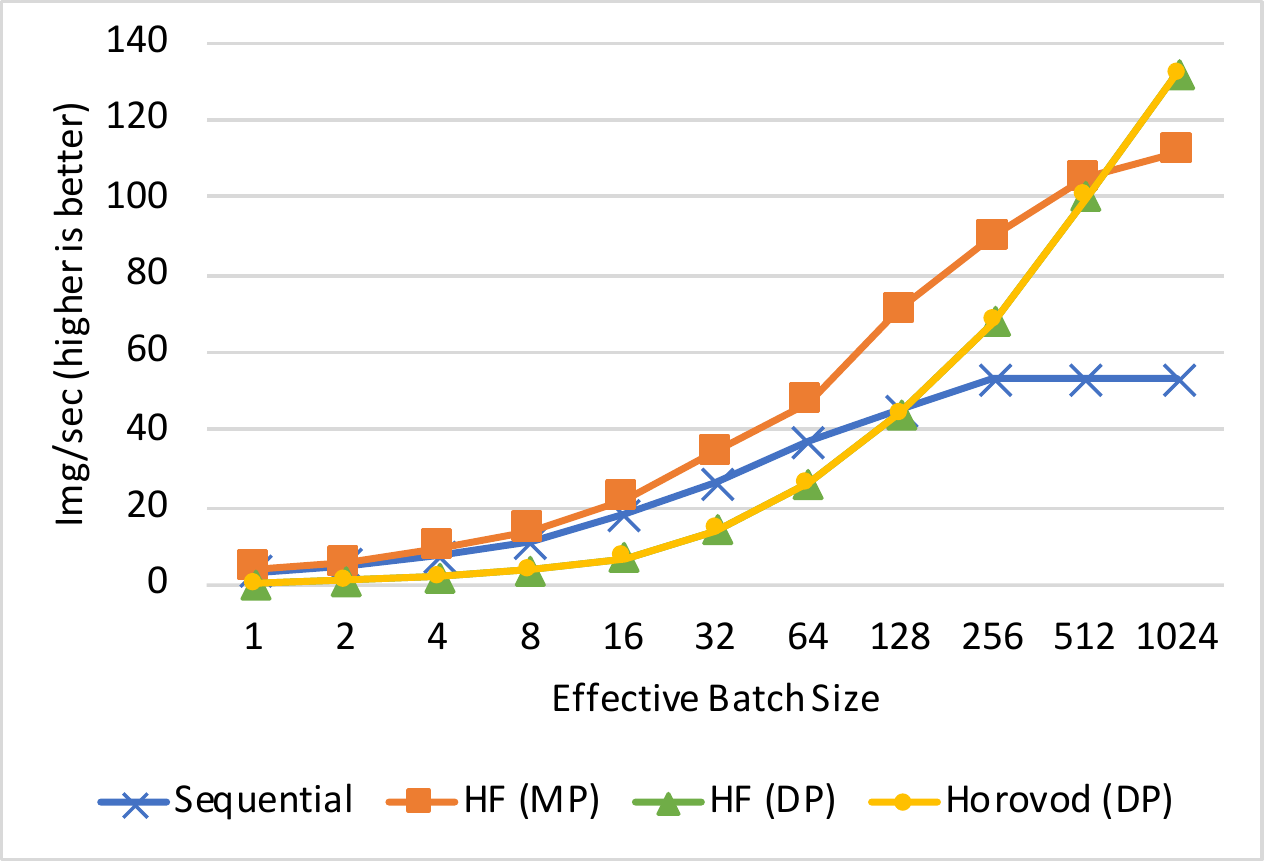}
{fig:vgg-single}
{ResNet-110-v1 up to 48 model-partitions}
{graphs/res110-seq-vs-mp-skylake-bs-one-node.pdf}
{fig:res110-single}
{HyPar-Flow's Model-Parallelism vs. Sequential/Data-Parallelism}

\mytwocolfig
{ResNet-110-v1 up to 64 model-partitions}
{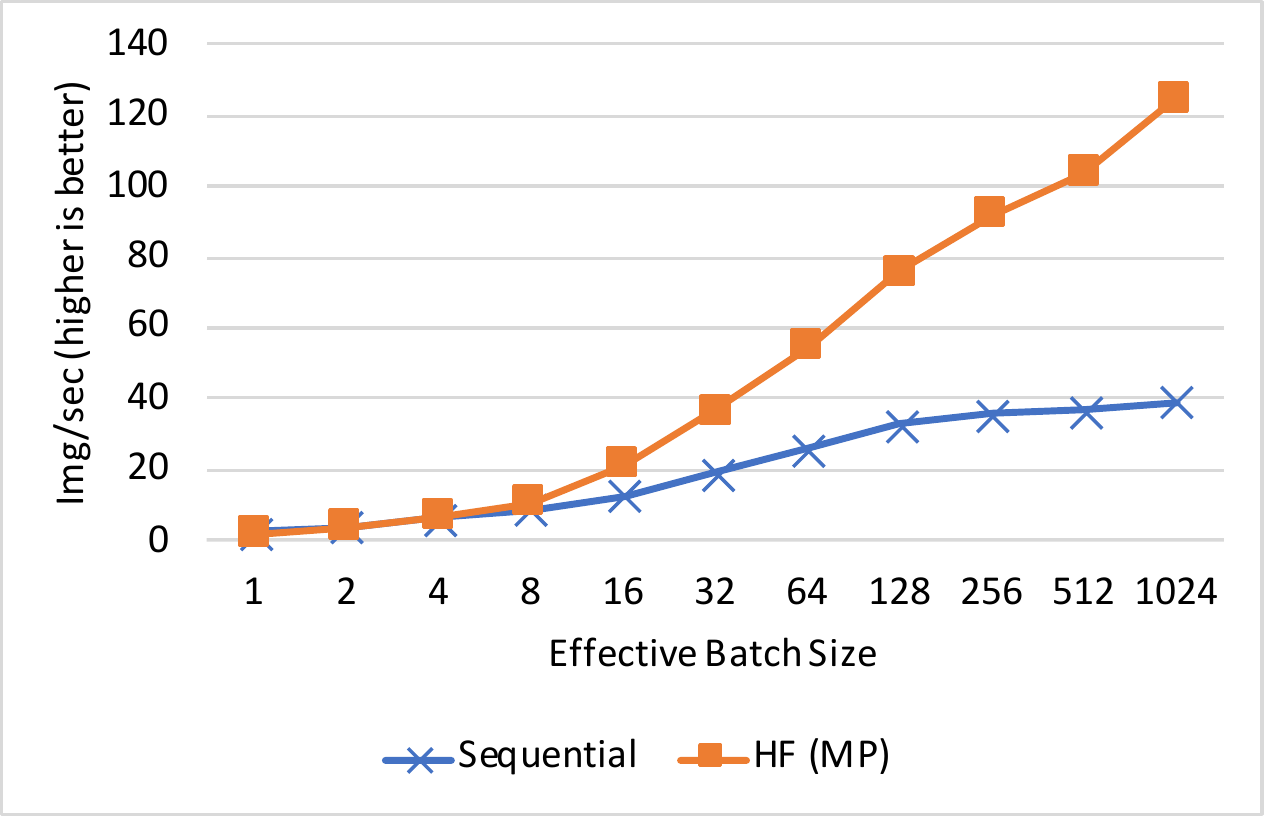}
{fig:epyc}
{ResNet-1001-v2 up to 48 model-partitions}
{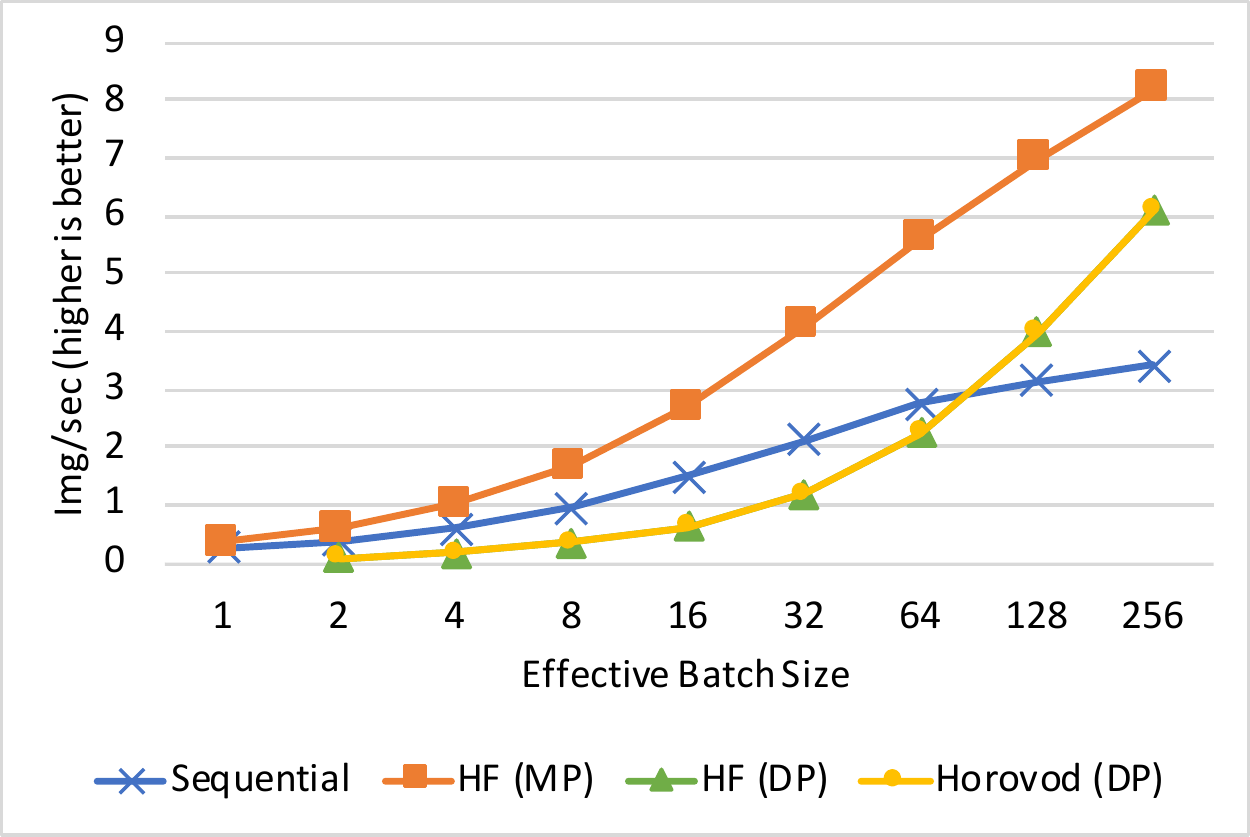}
{fig:res1k-single}
{HyPar-Flow's Model-Parallelism vs. Sequential/Data-Parallelism}

\MySubsection{Model Parallelism on Two Nodes}
\label{sec:multi-1}

To save space, two-node results are presented for VGG-16 and ResNet-1001-v2 only. Figure~\ref{fig:vgg-two-node} shows the performance trends for VGG-16 training across two nodes. As mentioned earlier, we are only able to achieve good performance with model-parallelism for up to 8 model-partitions for the 16 layers of VGG-16. We also perform experiments for 16 model-partitions but observe performance degradation. This is expected because of the lesser computation per partition and greater communication overhead in this scenario. We scale ResNet-1001-v2 on two nodes using 96 model-partitions in the model-parallelism-only configuration on Stampede2. The result is presented in Figure~\ref{fig:res1k-two-node}. We observe that model-parallel HF (MP) training provides 1.6$\times$ speedup (at BS=256) over HF (DP) and Horovod (DP). On the other hand, a data-parallel-only configuration is not able to achieve good performance for ResNet-1001 due to significant communication (allreduce) overhead during gradient aggregation.

\mytwocolfig
{VGG-16: MP Good for Small BS vs. DP Good for Large BS (8 model-partitions)}
{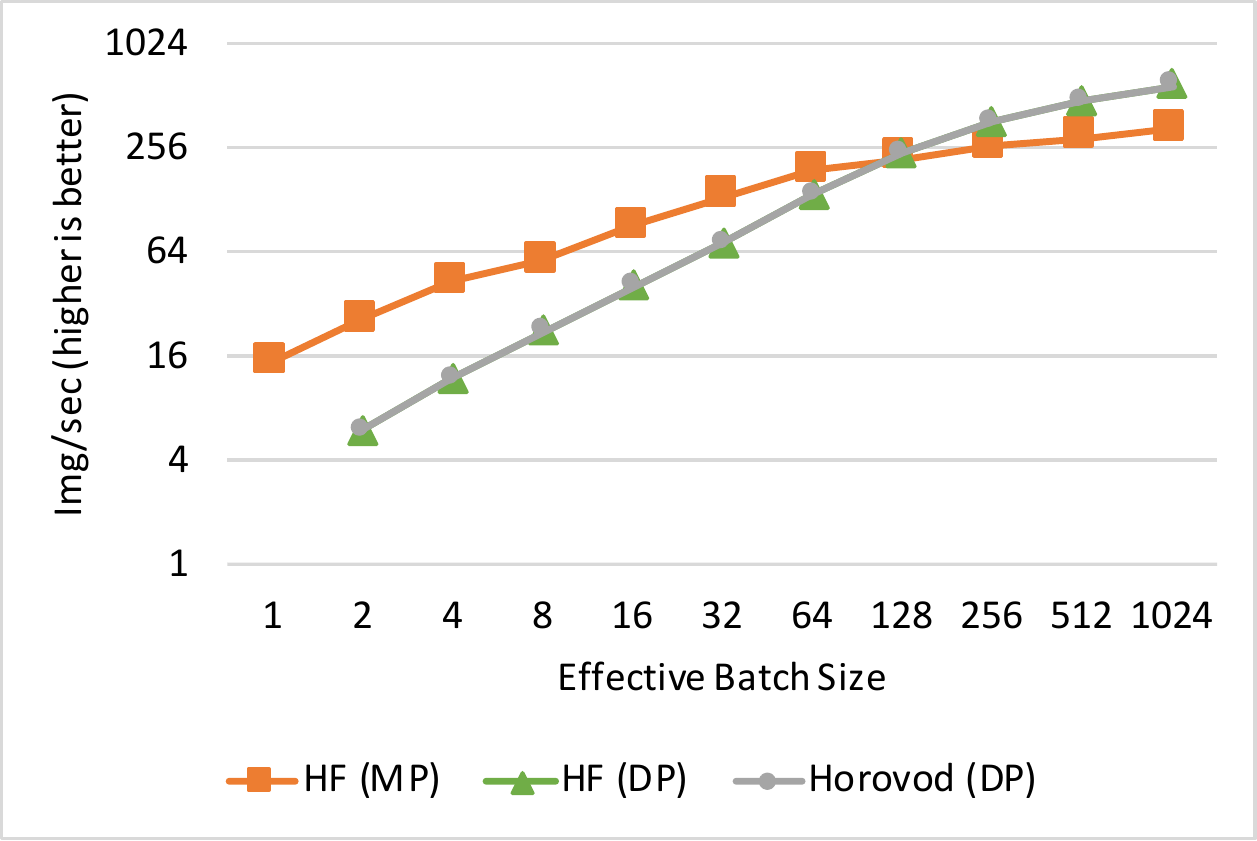}
{fig:vgg-two-node}
{ResNet-1001-v2: MP Good for All BS (up to 96 model-partitions).}
{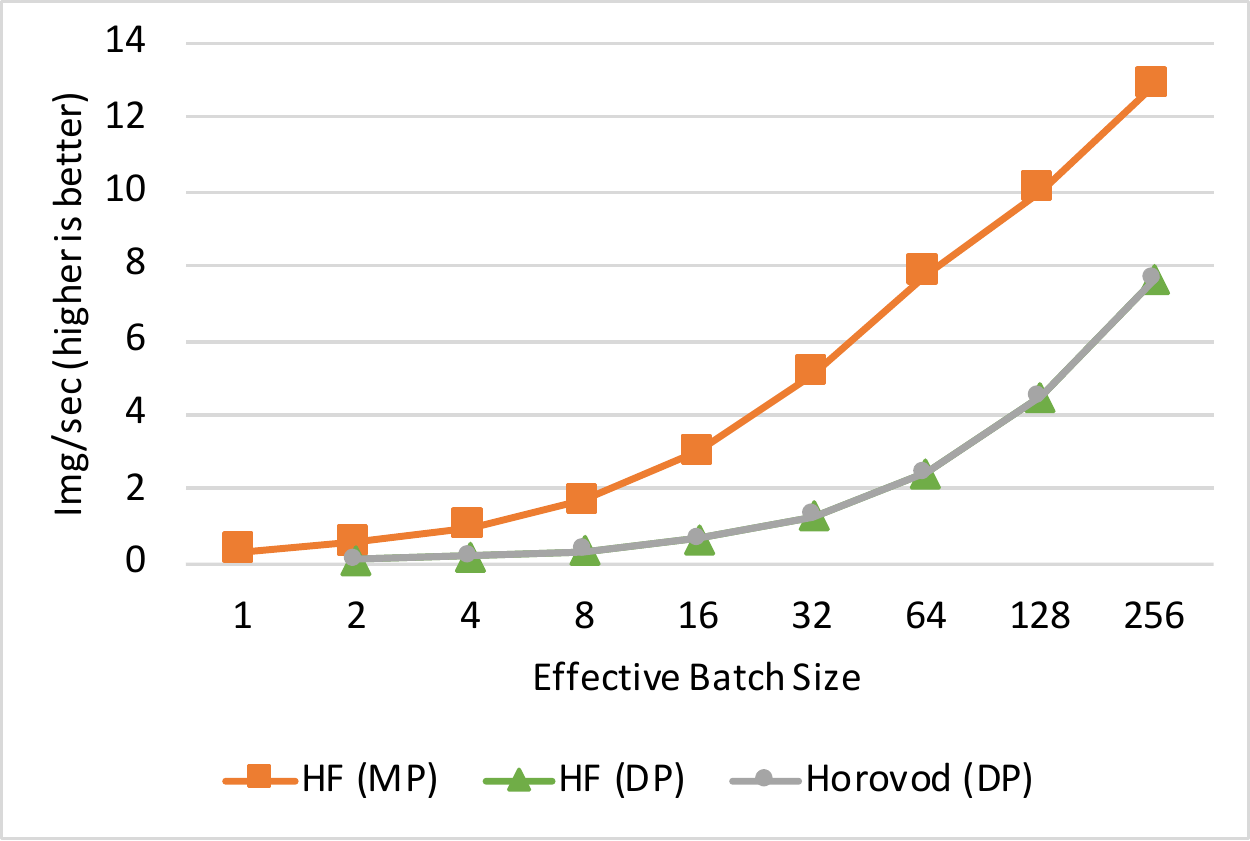}
{fig:res1k-two-node}
{HyPar-Flow Model-Parallelism across Two Nodes}

\MySubsection{Hybrid Parallelism at Scale: Up to 28,762 cores on 512 nodes}
\label{sec:multi-2}

The most comprehensive coverage of HyPar-Flow's flexibility, performance, and scalability are presented in Figure~\ref{fig:hybrid-stampede}. The figure shows performance for various combinations of hybrid-parallel training of ResNet-1001-v2 on 128 Stampede2 nodes. The figure has three dimensions: 1) the number of nodes on the X-axis, 2) Performance (Img/sec) on Y-axis, and 3) Batch Size using the diameter of the circles. The key takeaway is that hybrid-parallelism offers the user to make trade-offs between high-throughput (Img/sec) and batch size. From an accuracy (convergence) standpoint, the goal is to keep the batch-size small so model updates are more frequent. However, larger batch-size delays synchronization and thus provides higher throughput (Img/sec). HyPar-Flow offers the flexibility to control these two goals via different configurations. For instance, the large blue circle with diagonal lines shows results for 128 nodes using 128 model-replicas where the model is split into 48 partitions on the single 48-core node. This leads to a batch-size of just 32,768, which is 2$\times$ smaller than the expected 65,536 if pure data-parallelism is used. It is worth noting that the performance of pure data-parallelism even with 2$\times$ larger batch-size will still be lesser than the hybrid-parallel case, i.e., 793 img/sec (=6.2$\times$128 -- considering ideal scaling for data-parallel case presented earlier in Figure~\ref{fig:res1k-single}) vs. 940 img/sec (observed value-- Figure~\ref{fig:hybrid-stampede}). This is a significant benefit of hybrid-parallel training, which is impossible with pure model and/or data parallelism. In addition to this, we also present the largest scale we know of for any model/hybrid-parallel study on the latest Frontera system. Figure~\ref{fig:hybrid-frontera}) shows near-ideal scaling on 512 Frontera nodes. Effectively, every single core out of the 28,762 cores on these 512 nodes is being utilized by HyPar-Flow. The ResNet-1001 model is split into 56 partitions as Frontera nodes have a dual-socket Cascade-Lake Xeon processor for a total of 56 cores/node. We run one model-replica per node with a batch size of 128. To get the best performance, pipeline stages were tuned and the best number was found to be 128.

\mytwocolfig
{128 Stampede2 nodes}
{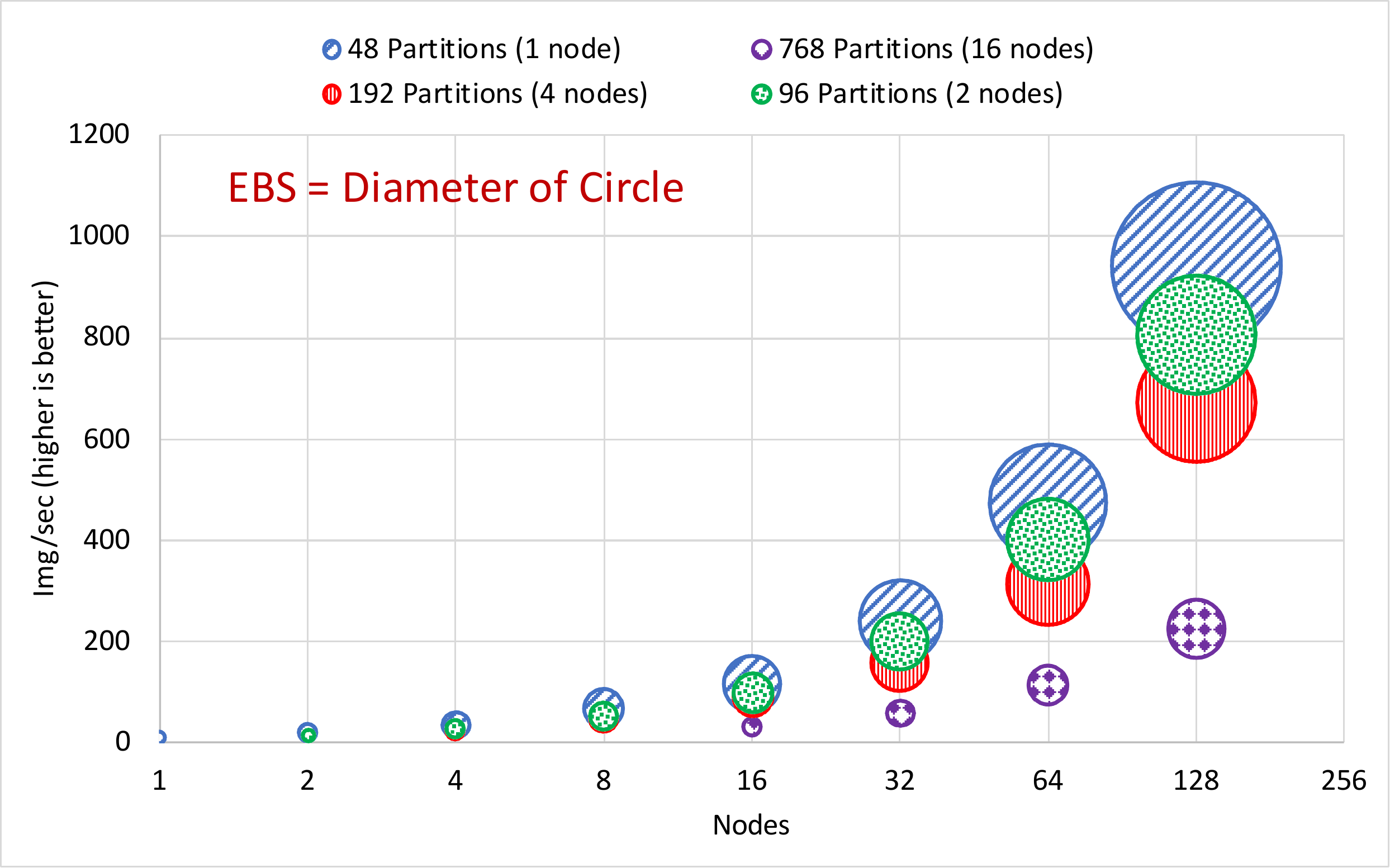}
{fig:hybrid-stampede}
{512 Frontera nodes}
{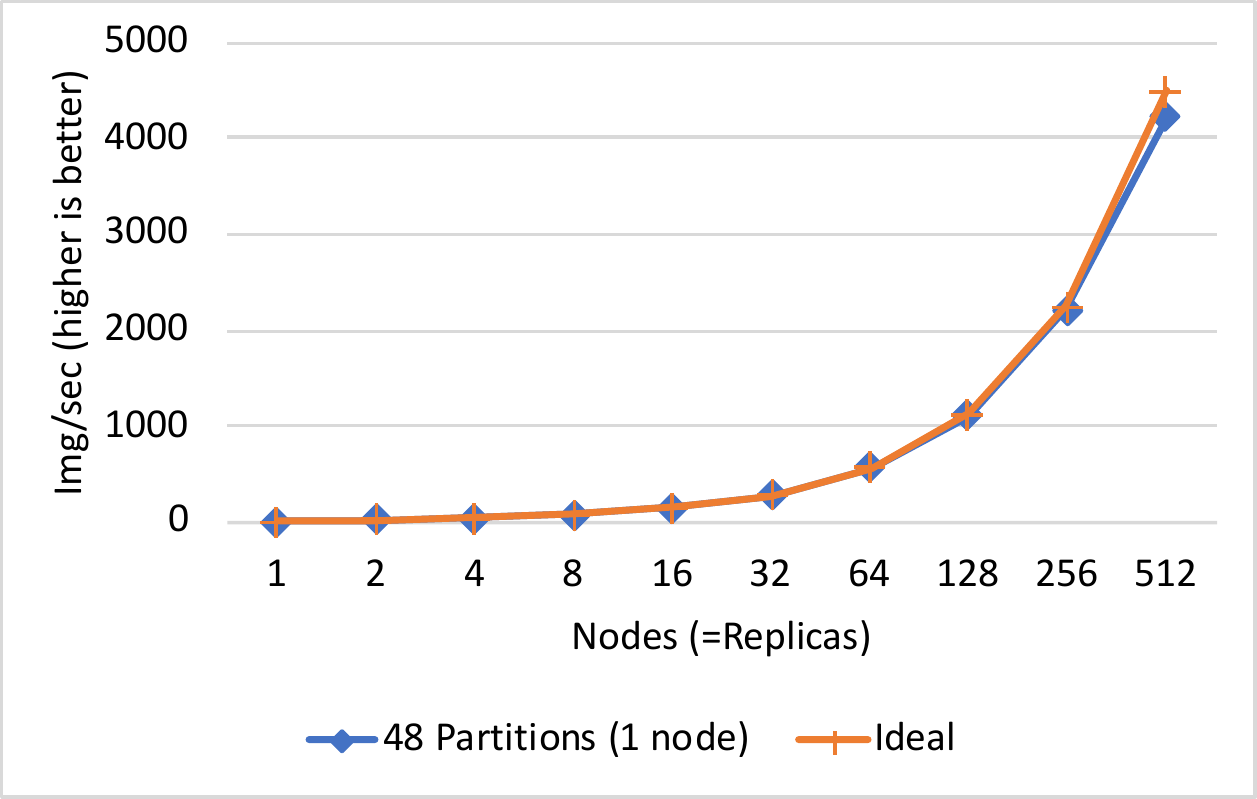}
{fig:hybrid-frontera}
{Hybrid-Parallelism at Scale: ResNet-1001-v2 on Stampede and Frontera with different batch sizes, number of replicas, and number of partitions}

\MySubsection{Next-generation Models: ResNet-5000?}

Today, designers develop models accounting for the restriction of memory consumption. However, with HyPar-Flow, this restriction no longer exists, and designers can come up with models with as many layers as needed to achieve the desired accuracy. To illustrate this, we present ResNet-5000, an experimental model with 5000 layers. ResNet-5000 is massive and requires a lot of memory so we were able to train it with a batch-size of 1 only. Beyond that, it is not trainable on any existing system. We stress-test HyPar-Flow to scale the training of ResNet-5000 to two nodes and were able to train for bigger batch sizes. We note that training ResNet-5000 and investigation of its accuracy and finding the right set of hyper-parameters is beyond the scope of this paper. The objective is to showcase HyPar-Flow's ability to deal with models that do not exist today.

\MySubsection{Discussion and Summary of Results}
\label{sec:insights}

Model and data-parallelism can be combined in a myriad of ways to realize hybrid-parallel training. E.g. model-parallelism on a single node with multiple cores with data-parallelism across nodes. There are non-trivial and model-dependent trade-offs involved when designing hybrid schemes. Model-parallelism and data-parallelism have different use cases; model-parallelism is beneficial when we have a large model, or we want to keep a small effective batch size for training. On the other hand, data-parallelism gives a near-linear scale-out on multiple nodes but it also increases batch size. In our experiments, we observe that single-node model-parallelism is better than single-node data-parallelism. Theoretically, the number of model-partitions can not be larger than the number of layers in the model; we can not have more than 110 partitions for ResNet-110. In practice, however, we observe that one layer per model-partition will not be used because it suffers from performance degradation. To conclude, HyPar-Flow's flexible hybrid-parallelism offers the best of both worlds; we can benefit from both model and data parallelism for the same model. We summarize the key observations below: 

\begin{itemize}
\vspace{-1.5ex}    
    \item Models like ResNet-110 offer better performance for model-parallelism on smaller batch sizes ($<$128). 
    
    \item Newer and very-deep models like ResNet-1001 benefit from model-parallelism for any batch size (Figure~\ref{fig:res1k-single}).
    
    \item HyPar-Flow's model-parallel training provides up to 3.2$\times$ speedup over sequential training and 1.6$\times$ speedup over data-parallel training (Figure~\ref{fig:epyc}).
    
    \item HyPar-Flow's hybrid-parallel training offers flexible configurations and provides excellent performance for ResNet-1001; 110$\times$ speedup over single-node training on 128 Stampede2 (Xeon Skylake) nodes (Figure~\ref{fig:hybrid-stampede}).
    
    \item HyPar-Flow's hybrid-parallel training is highly scalable; we scale ResNet-1001 to 512 Frontera nodes (28,762 cores) as shown in Figure~\ref{fig:hybrid-frontera}.
\end{itemize}

\vspace{-2.0ex}
\MySection{Conclusion}

Deep Learning workloads are going through a rapid change as newer models and larger, more diverse datasets are being developed. This has led to an explosion of software frameworks like TensorFlow and approaches like data and model-parallelism to deal with ever-increasing workloads. In this paper, we explored a new approach to train state-of-the-art DNNs and presented HyPar-Flow: a unified framework that enables user-transparent and parallel training of TensorFlow models using multiple parallelization strategies. HyPar-Flow does not enforce any specific paradigm. It allows the programmers to experiment with different parallelization strategies without requiring any changes to the model definition and without the need for any system-specific parallel training code. Instead, HyPar-Flow Trainer and Communication Engine take care of assigning the partitions to different processes and performing inter-partition and inter-replica communication efficiently. For ResNet-1001 training using HyPar-Flow, we were able to achieve excellent speedups: up to 1.6$\times$ over data-parallel training, up to 110$\times$ over single-node training on 128 Stampede2 nodes, and up to 481$\times$ over single-node on 512 Frontera nodes. We also tested the ability of HyPar-Flow to train very large experimental models like ResNet-5000, which consists of 5,000 layers. We believe that this study paves new ways to design models. We plan to publicly release the HyPar-Flow system so that the community can use it to develop and train next-generation models on large-scale HPC systems.

\vspace{-2.5ex}
\bibliographystyle{splncs04}

\end{document}